\documentclass[12pt]{iopart}

\usepackage{graphicx,bm}
\usepackage{amssymb}

\begin{document}

\title{Magnons versus electrons in thermal spin transport through metallic interfaces}


{\tiny }
\author{M~Beens}

\address{Institute for Theoretical Physics, Utrecht
	University, Leuvenlaan 4, 3584 CE Utrecht, The Netherlands}

\author{J~P~Heremans}

\address{Department of Mechanical and Aerospace Engineering, The Ohio State University,
Columbus, Ohio 43210, United States}
\address{Department of Physics, The Ohio State University, Columbus, Ohio 43210, United States}
\address{Department of Materials Science Engineering, The Ohio State University, Columbus, Ohio, 43210, United States}

\author{Yaroslav Tserkovnyak}
\address{Department of Physics and Astronomy, University of California, Los Angeles, California 90095, USA}

\author{R~A~Duine}

\address{Institute for Theoretical Physics, Utrecht
	University, Leuvenlaan 4, 3584 CE Utrecht, The Netherlands}

\address{Department of Applied Physics, Eindhoven University of Technology, P.O. Box 513, 5600 MB Eindhoven, The Netherlands}

\begin{abstract} We develop a theory for spin transport in magnetic metals that treats the contribution of magnons and electrons on equal footing. As an application we consider thermally-driven spin injection across an interface between a magnetic metal and a normal metal, i.e., the spin-dependent Seebeck effect. We show that the ratio between magnonic and electronic contribution scales as $\sqrt{T/T_C}T_F/T_C$, with the Fermi temperature $T_F$ and the Curie temperature $T_C$. Since, typically, $T_C \ll T_F$, the magnonic contribution may dominate the thermal spin injection, even though the interface is more transparent for electronic spin current. 
\end{abstract}

\section{Introduction}

Over the past decade the interplay between spin, charge and heat currents has attracted considerable attention and has led to the field dubbed ``spin caloritronics" \cite{bauer2012}. Central to this field is the thermally-driven injection of spin current from a magnetic material into a normal metal across an interface between them. In case of the spin Seebeck effect \cite{uchida2010,PhysRevLett.106.186601} the magnetic material is an insulator, typically Yttrium Iron Garnett, while the normal metal is typically Pt. The injected spin current then manifests itself as a voltage across the Pt that is transverse to the interface normal and arises as a result of the inverse spin Hall effect in the Pt. The term  {\it spin-dependent} Seebeck effect is nowadays restricted to the situation in which the magnetic material is a metal. This latter effect was first observed in a non-local geometry \cite{slachter2010} using permalloy and Cu as the respective magnetic and normal metals.

The models for the spin Seebeck effect invoke magnons in the magnetic insulator as the carriers of the spin current \cite{PhysRevB.81.214418,PhysRevB.88.064408}. At the interface between the insulator and the normal metal, the magnonic spin current that flows in response to a temperature gradient is converted into electronic spin current in the normal metal by interfacial spin-flip scattering processes. The existing models for the spin-dependent Seebeck effect \cite{bauer2012,slachter2010}, on the other hand, are spin-dependent drift-diffusion models for the electrons in the metallic ferromagnet and the normal metal, in which the interface can be taken to be essentially transparent for the electrons (as compared to the pertinent diffusive contributions from the bulk). Magnons in the magnetic metal are, in these models, neglected completely. A priori, there is no reason to discard the magnons as carriers of spin currents in magnetic metals. For example, the magnon-drag thermopower \cite{PhysRevLett.18.395}, i.e.,  the contribution to the charge Seebeck effect due to thermally driven magnons that drag along electrons, has been shown to dominate the thermopower in Fe, Ni, and Co, over a wide range of temperatures \cite{PhysRevB.94.144407}.  

In this special-issue contribution, we develop a theory for spin transport in metallic ferromagnets that treats magnonic and electronic spin currents on equal footing. While electronic spin currents dominate the spin transport when it is driven by an electric field, we find that for thermally driven spin transport the magnonic contribution cannot be neglected and may, in fact, be larger than the electronic one. This is because the magnitude of the latter is governed by the dimensionless ratio $T/T_F$, with the $T$ the temperature and $T_F$ the Fermi temperature. The thermally-driven magnonic spin current, on the other hand, is determined by $(T/T_C)^{3/2}$ with $T_C$ the Curie temperature. Since $T_F$ is at least one order of magnitude larger than $T_C$ for the most common metallic ferromagnets, the magnonic spin current may overwhelm the electronic contribution in the bulk in situations where the spin current is driven by a thermal gradient. As a result, one would naively expect that the magnonic contribution dominates the electronic contribution to the thermal spin injection from a magnetic to a nonmagnetic metal. For thermal spin injection the magnonic contribution is, however, relatively diminished somewhat because the interface between common ferromagnetic and normal metals is less transparent for magnonic spin current as compared to electronic spin current. 

The remainder of this work is organized as follows. In the next section, we develop our theory for spin transport in ferromagnetic metals that includes both electronic spin accumulation and magnon chemical potential, and estimate the various coefficients. In Sec.~\ref{sec:appl} we consider as an application thermal spin injection into a normal metal and discuss it in terms of an equivalent circuit. We end in Sec.~\ref{sec:discconcloutl} with a short discussion and outlook. 
\begin{figure}
	\includegraphics[width=15cm]{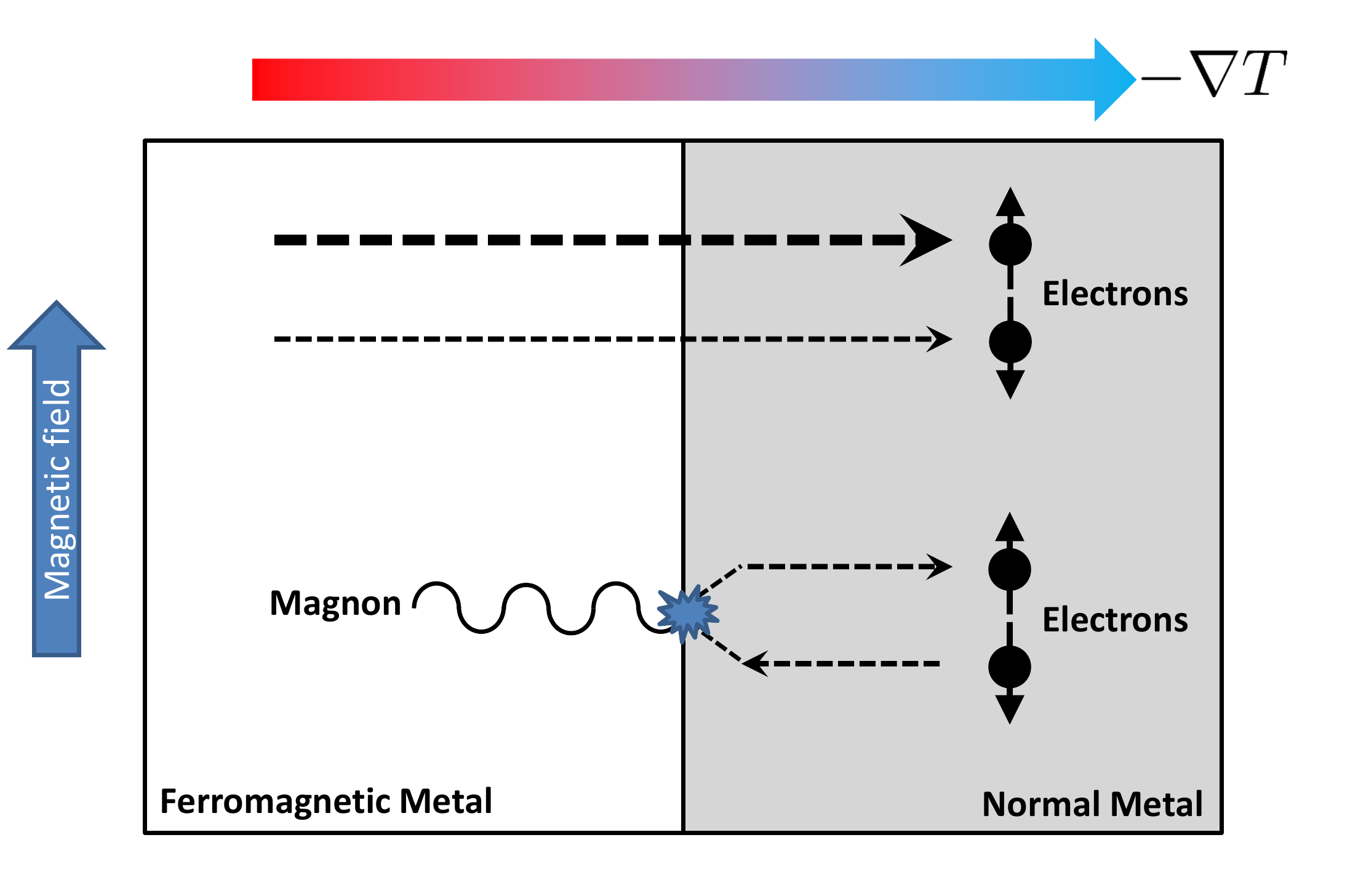} 
	\caption{The set-up that we consider in this article. A temperature gradient drives spin transport across an interface between a ferromagnetic and normal metal. The resultant spin injection from ferromagnet into the normal metal occurs via electrons that are spin polarized in the ferromagnet and that flow from the magnet into the normal metal, in parallel with processes in which a magnon induces a spin current in the normal metal via interfacial electron-magnon scattering.}
	\label{fig:system}
\end{figure}

\section{Theory}
\label{sec:gentheory}

The set-up we consider is sketched in Fig.~\ref{fig:system}. We consider a ferromagnetic metal with a sufficiently large magnetic field that is applied in the $\hat z$-direction, so that the spin density is saturated in the $-\hat z$-direction. Since we will in the following mostly consider thermal magnons at room temperature, we ignore their ellipticity. It then follows that magnons carry spin angular momentum $+\hbar$. The ferromagnet is interfaced with a normal metal, and we are mostly interested in the spin current that is injected from the ferromagnet into the normal metal as a result of an applied temperature gradient $\nabla T$. We consider two processes that contribute to this spin injection. First, there is a thermally-excited electron spin current that traverses the interface. Second, there is a thermally-excited magnon spin current that is converted at the interface into an electronic spin current in the normal metal by interfacial electron-magnon spin-flip scattering. In this section we develop a simple theory that takes both processes into account on equal footing, and give estimates of the various parameters that enter our theory. In the development of our theory we combine the drift-diffusion theory for the electronic contribution to thermal spin transport \cite{PhysRevB.84.174408,PhysRevB.35.4959} with the theory for spin transport in insulator-metal hybrids developed by several of us \cite{PhysRevLett.116.117201,PhysRevB.94.014412}. Below, we do not explicitly include references in case the results can be found in any of these works. Before introducing our theory, however, we discuss some simplifying assumptions. 

\subsection{Preliminary remarks}
\label{subsec:prelremarks}
Our general starting point is to treat the magnons, phonons and electrons in the ferromagnetic metal as internally equilibrated subsystems that may exchange heat and spin. The heat exchange is driven by differences between the magnon ($T_m$), phonon ($T_p$) and electron ($T_e$) temperatures. Similarly, the exchange of spin is driven by the difference between the magnon chemical potential ($\mu_m$) and the electron spin accumulation $\mu_s=\mu_\uparrow - \mu_\downarrow$, where $\mu_\uparrow$ and $\mu_\downarrow$ are the chemical potentials of the electrons with spin projection along and against the external magnetic field, respectively. While writing down a complete phenomenological theory that takes into account all processes of spin and heat exchange between magnons, phonons and electrons is in principle straightforward, such a theory is somewhat untractable because of the amount of free parameters, given that ``off-diagonal" processes --- e.g. spin exchange driven be temperature differences --- also need to be taken into account. As our goal is to develop a simple phenomenological theory that treats the magnon and electron spin transport on equal footing and to show that the contribution of magnons is not negligible, we limit ourselves to the situation that all temperatures are equal, $T_e=T_p=T_m$, and assume our system is described by the magnon chemical potential, electron spin accumulation, and one temperature. We assume that anharmonicities lead to fast phonon number decay, so that the phonon chemical potential $\mu_p$ is taken to be zero always. 

\begin{figure}
	\includegraphics[width=15cm]{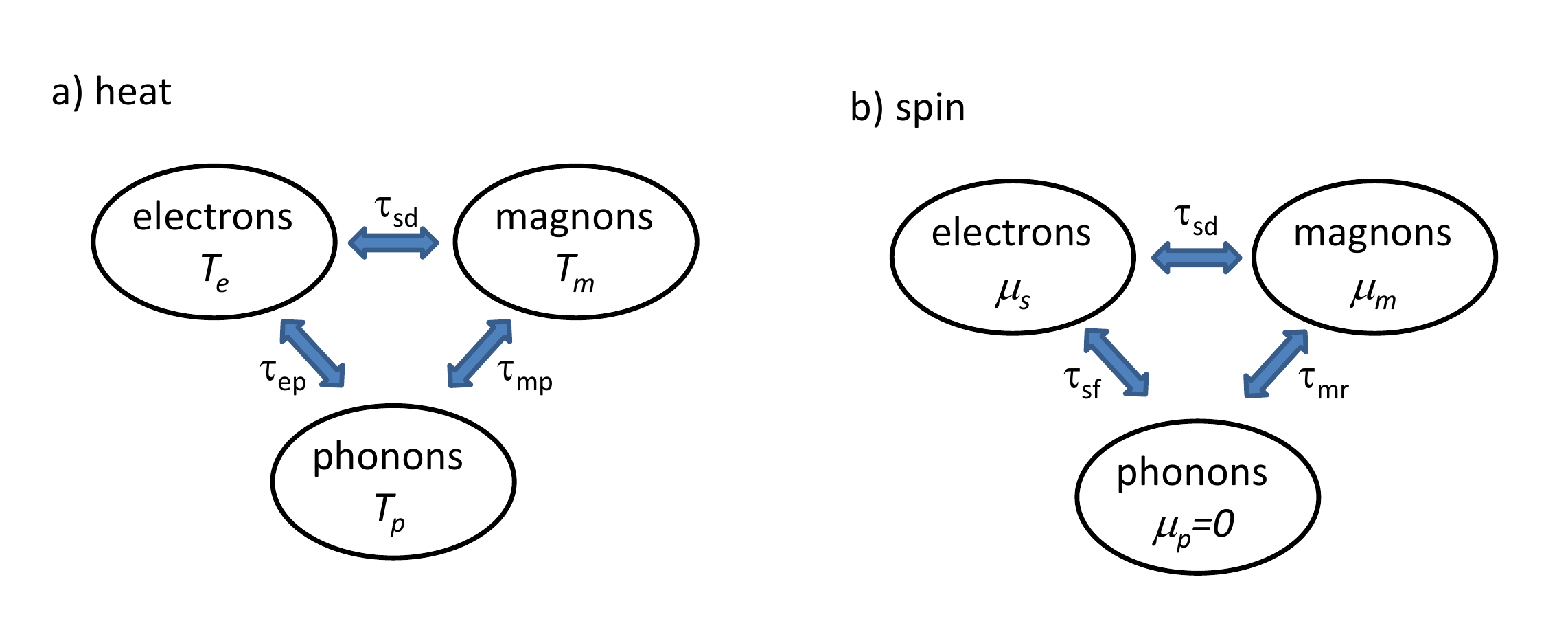} 
	\caption{Schematic that indicates the time scales for a) the heat exchange between electronic, magnonic and phononic reservoirs and b) the exchange of spin.}
	\label{fig:exchangeschematic}
\end{figure}

These restrictions follow from assumptions on the hierarchy of time scales that characterize the various heat and spin exchange processes. In Fig.~\ref{fig:exchangeschematic} we indicate these relaxation times for both heat (a) and spin (b) exchange. (See Table~\ref{tab:relaxationtimes} for a list of all relaxation times used in this article.)  Here, the interaction between electrons and magnons is assumed to be dominated by $s$-$d$-scattering --- an electron spin-flip accompanied by absorption or emission of a magnon --- so that the corresponding time scale is labeled $\tau_{\rm sd}$. This process governs both heat and spin exchange between magnons and electrons. The time scale for electron-phonon scattering is indicated by $\tau_{\rm ep}$. Furthermore, $\tau_{\rm mp}$ is the time scale for all magnon-phonon collisions, while $\tau_{\rm mr} \geq \tau_{\rm mp}$ is due to magnon non-conserving magnon-phonon collisions only.  Similarly, $\tau_{\rm sf} \geq \tau_{\rm ep}$ is the time scale for electron-phonon spin-flip scattering. In later estimates we will assume a contribution due to disorder to $\tau_{\rm sf}$ and will thus take it to approach a constant as $T \to 0$. We assume that $\tau_{\rm ep}$ is the smallest time scale, so that $T_e=T_p$. We, furthermore, assume that $\tau_{\rm mp} \ll \tau_{\rm sd} $, so that we have that $T_p=T_m(=T_e) \equiv T$, while the difference between the electron spin accumulation and magnon chemical potential needs to be taken into account. In the next subsection we provide estimates for Fe at room temperature to underpin some of these assumptions. We remark that the hierarchy of time scales implies a hierarchy of length scales, i.e., describing the spin transport with magnon chemical potential, spin accumulation, and one temperature is --- within the assumptions on time scales mentioned here --- only applicable for sufficiently long lengths.  

\subsection{Estimates}
Let us now make some estimates for pure Fe. We deduce the electronic transport relaxation time $\tau_{{\rm tr},e}$ from the electronic conductivity $\sigma \sim 10^7$ S$/$m of Fe  \cite{PhysRevB.94.144407}, so that we find --- using the Drude formula --- that $\tau_{{\rm tr},e} \sim m \sigma/n_e e^2$. Taking for the effective mass $m$ the bare electronic mass, and using an electron density of $n_e \sim (0.3~{\rm nm})^{-3}$, we find that $\tau_{{\rm tr},e} \sim 0.01-0.1$~ps. This time scale includes all electronic momentum non-conserving scattering events, and, in particular, electron-phonon scattering and spin-conserving electron-electron $s$-$d$-scattering. While the latter may dominate, we assume here that we may estimate $\tau_{\rm ep} \sim \tau_{{\rm tr},e}$. 

The term $\tau_{\rm sd}$ is used here to denote the interaction between electrons and magnons, an electron spin-flip accompanied by the absorption or emission of a magnon. To estimate this scattering time we follow Ref.~\cite{PhysRevB.92.180412}. We then have that $\tau_{\rm sd} \sim \hbar/\eta k_B T$, where $\eta$ is itself a function of temperature and which we estimate $\eta \sim 0.01$ at a temperature that is a fraction of $T_C$ \cite{PhysRevB.92.180412}. We then find that $\tau_{\rm sd} \sim 10$~ps $\gg\tau_{\rm ep}$ at room temperature. 

The scattering time for all magnon-phonon collisions is $\tau_{\rm mp}$ and includes both magnon-conserving and magnon-non-conserving processes. While there is to the best of our knowledge no direct measurement of this time scale, Ref.~\cite{PhysRevB.94.014412} estimates it to be on the order of several ps. Hence, we may suppose that  $\tau_{\rm sd} \gtrsim \tau_{\rm mp}$, although the opposite limit can also be realized. For simplicity, we restrict ourselves to the former regime.

\begin{table}
	\caption{\label{tab:relaxationtimes} Relaxation times}
	\footnotesize\rm
	\begin{tabular*}{\textwidth}{@{}l*{15}{@{\extracolsep{0pt plus12pt}}l}}
		\br
		Symbol & Meaning \\
		\mr
		$\tau_{\rm ep}$ & Electron-phonon scattering time \\
		$\tau_{\rm sd}$ & Electron-magnon scattering time \\
	    $\tau_{\rm mp}$ & Magnon-phonon scattering time \\
		$\tau_{\rm sf}$ & Electron spin-flip time \\
	    $\tau_{\rm mr}$ & Magnon-number non-conserving magnon-phonon scattering time\\
		$\tau_{{\rm tr},e}$ & Electron-momentum relaxation time\\
		$\tau_{{\rm tr},m}$ & Magnon-momentum relaxation time\\
		$\tau_{\rm em}, \tau_{\rm me}$ & Electron-magnon momentum transfer time\\
		\br
	\end{tabular*}
\end{table}

\subsection{Theory for the bulk spin transport in the ferromagnet}
\label{subsec:phentheory}

Within the assumptions discussed so far, we develop a theory based on conservation of spin. The density of spin-up electrons is $n_\uparrow$, while that of spin-down electrons is $n_\downarrow$. The density of magnons is $n_m$. The three resultant continuity equations are
\begin{eqnarray}
\label{eq:conteqs}
\frac{\partial n_\uparrow}{\partial t} + \nabla \cdot {\bf j}_\uparrow &=& - \frac{\nu_\uparrow}{\tau_{{\rm sf},\uparrow}} \mu_s - \frac{\nu_\uparrow}{\tau_{{\rm sd},\uparrow}} \left(\mu_s - \mu_m \right)~, \nonumber \\
\frac{\partial n_\downarrow}{\partial t} + \nabla \cdot {\bf j}_\downarrow &=&  +\frac{\nu_\downarrow}{\tau_{{\rm sf},\downarrow}} \mu_s + \frac{\nu_\downarrow}{\tau_{{\rm sd},\downarrow}} \left(\mu_s - \mu_m \right)~, \nonumber \\
\frac{\partial n_m}{\partial t} + \nabla \cdot {\bf j}_m &=&  -\frac{\chi_m}{\tau_{\rm mr}} \mu_m + \frac{\chi_m}{\tau_{\rm sd}} \left(\mu_s - \mu_m \right)~.
\end{eqnarray}
Here $\nu_\uparrow$ and $\nu_\downarrow$ are the electronic density of states for up and down electrons, respectively, and $\chi_m$ is the magnon susceptibility. Note that the electron spin-flip time, and $s$-$d$-scattering times are in principle spin dependent, which we ignored in our discussion in the previous subsection and in Fig.~\ref{fig:exchangeschematic}. Spin conservation of the $s$-$d$-interactions implies
\[
   \frac{\chi_m}{\tau_{\rm sd}} = \frac{1}{2} \left(\frac{\nu_\uparrow}{\tau_{{\rm sd},\uparrow}}+ \frac{\nu_\downarrow}{\tau_{{\rm sd},\downarrow}} \right)~,
\] 
while charge conservation yields
\[
  \frac{\nu_\uparrow}{\tau_{{\rm sf},\uparrow}}= \frac{\nu_\downarrow}{\tau_{{\rm sf},\downarrow}}~,
\]
and
\[
\frac{\nu_\uparrow}{\tau_{{\rm sd},\uparrow}}= \frac{\nu_\downarrow}{\tau_{{\rm sd},\downarrow}}~.
\]

The spin-resolved charge currents $-e {\bf j}_\alpha$ --- with the index $\alpha \in \{\uparrow, \downarrow\}$ while the respective number $\alpha \in \{+,-\}$ --- obey the linear-response expressions
\begin{equation}
\label{eq:electronspincurrent}
 - e {\bf j}_\alpha = \sigma_\alpha \frac{\nabla \mu_\alpha}{e} - \sigma_\alpha S_\alpha \nabla T + \sigma_{{\rm md},\alpha}  \frac{\nabla \mu_m}{e}~,
\end{equation}
where $\sigma_\alpha$ are the spin-dependent electron conductivities and where $e$ is minus the elecrton charge. The magnon-drag conductivities $\sigma_{{\rm md},\alpha}$ stem from frictional drag between electrons and magnons. The spin-dependent Seebeck coefficients are denoted by $S_\alpha$, and also include a contribution due to magnon drag that does not appear explicity since we have taken $T_e=T_m=T$. We neglect drag between the two spin projections of the electron as such interaction effects vanish quadratically with $T/T_F$. 

The magnon spin current is given by
\begin{eqnarray}
\label{eq:magnonspincurrent}
\hbar {\bf j}_m &=&  -\sigma_m \frac{\nabla \mu_m}{\hbar} - L \frac{\nabla T}{T} - \frac{\hbar}{e^2} \left(\sigma_{{\rm md},\uparrow} +\sigma_{{\rm md},\downarrow} \right) \nabla\mu_e \nonumber \\
&&- \frac{\hbar}{2 e^2} \left(\sigma_{{\rm md},\uparrow} -\sigma_{{\rm md},\downarrow} \right) \nabla\mu_s~,
\end{eqnarray}
with the magnon conductivity $\sigma_m$ and the bulk spin Seebeck coefficient $L$ (that latter also contains the magnon-drag contribution), and is driven by gradients in the magnon chemical potential, temperature, charge accumulation $\mu_e=(\mu_\uparrow+\mu_\downarrow)/2$, and spin accumulation, corresponding to the four respective terms on the right-hand side of the above equation. 

Assuming a steady state and constant temperature gradients, we find --- in addition to charge conservation, $\nabla \cdot ({\bf j}_\uparrow+{\bf j}_\downarrow)=0$ --- from the continuity equations and the expresions for the currents that
\begin{eqnarray}
\label{eq:steadystateeqschempots}
&&\nabla^2 \mu_s = \frac{\mu_s}{\ell_{\rm sf}^2} + \frac{1}{\ell_{\rm sd}^2} \left[\mu_s - \mu_m \right]~, \nonumber \\
&&\nabla^2 \mu_m 
 = \frac{\mu_m}{\ell_{\rm mr}^2} +  \frac{\beta}{\ell_{\rm sd}^2} \left[\mu_m - \mu_s \right]~,
\end{eqnarray}
where we assumed that $\sigma_{\rm md,\alpha} \ll \sigma_m$, as estimated below, and neglected the magnon-drag conductivities. 
These equations involve the spin-flip relaxation length that governs decay of the electron spin accumulation
\begin{equation}
\frac{1}{\ell_{\rm sf}^2} = \frac{e^2\nu_\uparrow}{\sigma_\uparrow \tau_\uparrow} +  \frac{e^2\nu_\downarrow}{\sigma_\downarrow \tau_\downarrow}~,
\end{equation}
the length scale $\ell_{\rm sd}$ for electron-magnon spin equilibration given by
\begin{equation}
\frac{1}{\ell_{\rm sd}^2} = \frac{e^2\nu_\uparrow}{\sigma_\uparrow \tau_{{\rm sd},\uparrow}} +  \frac{e^2\nu_\downarrow}{\sigma_\downarrow \tau_{{\rm sd},\downarrow}}~,
\end{equation}
and the magnon relaxation length that governs relaxation of the magnon chemical potential
\begin{equation}
\frac{1}{\ell_{\rm mr}^2} =\frac{\chi_m \hbar^2}{\sigma_m \tau_{\rm mr}}~.
\end{equation}
The dimensionless constant $\beta=\hbar^2 \sigma_\uparrow\sigma_\downarrow /e^2(\sigma_\uparrow+\sigma_\downarrow)\sigma_m$ characterizes the electron conductivity relative to magnon one. Note that we have used the restrictions on the various time scales set by charge conservation and spin-conservation in the $s$-$d$-interactions in arriving at the expressions for the above length scales. 

We end this subsection by some remarks on how to extend the theory presented here beyond the simplifying assumptions that we made: The inclusion of separate temperatures for magnons, electrons, and phonons, would require one to include the continuity equations for the energy densities of magnons, electrons, and phonons, and the corresponding energy currents. The various continuity equations would have to include relaxation terms that correspond to the exchange processes depicted in Fig.~\ref{fig:exchangeschematic} as well as cross-relaxation terms. Finally, the six currents are driven by the various forces, giving rise to many more transport coefficients beyond the ones discussed so far.

\subsection{Remarks on magnon drag}
The magnon drag is discussed further using a simple hydrodynamic model \cite{PhysRevB.94.144407} that is known to give similar results as models that include spin-relaxation processes \cite{doi:10.1063/1.3672207, 0295-5075-115-5-57004}. To extract the magnon-drag conductivity from this model, we need only to include an electric field as the driving force. Ignoring, for the sake of simplicity, the spin dependence of the various quantities we have for the electron drift velocity ${\bf v}_e$ and the magnon drift velocity ${\bf v}_m$ the equations of motion
\begin{eqnarray}
\frac{d{\bf v}_e}{dt} = \frac{e {\bf E}}{m} - \frac{{\bf v}_e}{\tau_{{\rm tr},e}} - \frac{\left({\bf v}_e -{\bf v}_m\right)}{\tau_{\rm em}} ~, \nonumber \\
\frac{d{\bf v}_m}{dt} =  - \frac{{\bf v}_m}{\tau_{{\rm tr},m}} - \frac{\left({\bf v}_e -{\bf v}_m\right)}{\tau_{\rm me}}~,
\end{eqnarray}
where ${\bf E}$ is the applied electric field. Both magnons and electrons are assumed to have parabolic dispersion, with effective masses $m$ and $M$, respectively. The magnon transport relaxation time is $\tau_{{\rm tr},m}$, whereas the time scales $\tau_{\rm em}$ and $\tau_{\rm me}$ characterize momentum transfer between electrons and magnons. Momentum conservation yields $n_e m/\tau_{\rm em} = n_m M/\tau_{\rm me}$. Solving the above equations in the steady state for the magnon drift velocity, and using that ${\bf j}_m = n_m {\bf v}_m$, one finds that ${\bf j}_m = \sigma_{\rm md} {\bf E}/e$,
with 
\begin{equation}
\label{eq:mdcond}
\sigma_{\rm md} = \frac{n_e e^2}{m} \left(\frac{\tau_{{\rm tr},e} \tau_{{\rm tr},m}}{\tau_{\rm me}}\right)
\left(\frac{1}{\frac{M\tau_{{\rm tr},e}}{m \tau_{\rm me}} + \frac{n_e \tau_{{\rm tr},m}}{n_m \tau_{\rm me}} +\frac{M}{m}}\right)~.
\end{equation}
Using for $m$ again the bare electron mass, we have that for metallic ferromagnets $M/m\sim 100$. For the magnon density we estimate $n_m \sim n_e \left(T/T_C \right)^{3/2}$. Furthermore, taking $\tau_{\rm me} \sim \tau_{\rm sd}\sim \tau_{{\rm tr},m}$ and using our previous estimates we find that $\tau_{{\rm tr},e} \ll \tau_{\rm me}$. These estimates imply that generically $\sigma_{\rm md} \ll \left(\sigma_\uparrow + \sigma_\downarrow \right), e^2\sigma_m/\hbar$ where we used the Drude formulae $\left(\sigma_\uparrow + \sigma_\downarrow \right) = n_e e^2 \tau_{{\rm tr},e}/m$ and $\sigma_m = n_m \hbar^2\tau_{{\rm tr},m}/M$. For $T \to 0$ we have from Eq.~(\ref{eq:mdcond}) that $\sigma_{\rm md} \sim \left(\sigma_\uparrow + \sigma_\downarrow \right) \left(T/T_C\right)^{3/2}$, so that the magnonic spin current that is dragged along by the electronic charge current ${\bf j}_e = -e ({\bf j}_\uparrow+{\bf j}_\downarrow)$ in a ferromagnetic metal is
\begin{equation}
\hbar {\bf j}_m \sim -\left( \frac{T}{T_C}\right)^{3/2}\frac{{\bf j}_e}{e}~.
\end{equation}
These estimates show that the magnon-drag contribution to the conductivities can be neglected while the magnon drag turns out to give an important contribution to the thermopower for some materials, like Fe \cite{PhysRevB.94.144407}. In the theory presented here this latter contribution is in principle included in the coefficients $L$ and $S_\alpha$. 

\subsection{Spin transport in the normal metal} 
The equations that describe the spin transport on the normal-metal side are similar to the ones for the ferromagnet, with the modifications that there are no magnons present, and that there is no spin-dependence of transport coefficients. To distinguish the various quantities on the normal-metal side from those in the ferromagnet, we will give them the superscript ``$N$". For example, the linear-response expression for the current in Eq.~(\ref{eq:electronspincurrent}) becomes
\begin{equation}
\label{eq:electronspincurrentnormal}
- e {\bf j}_\alpha^N = \sigma^N \frac{\nabla \mu^N_\alpha}{e} - \sigma^N S^N \nabla T~.
\end{equation}
Also note that the magnon-drag contribution does not exist in the normal metal.

\subsection{Interfacial spin transport} As we are interested in thermal spin injection from a ferromagnet into a normal metal, we have to complement our bulk expressions for the spin current with an expression of the spin current through the interface. 
We assume that the interface is transparent for electrons, which implies that the electron spin accumulation is continuous at the interface. Ignoring loss of spin at the interface, we have, furthermore, for the currents through the interface that
\begin{equation}
\label{eq:bcspincurrents}
\left. \frac{1}{2} \left({j}_\uparrow -{j}_\downarrow \right) + {j}_m \right|_{\rm int} = \left. \frac{1}{2} \left({j}^N_\uparrow -{ j}^N_\downarrow \right) \right|_{\rm int}~.
\end{equation}
The magnon spin current at the interface is limited by interfacial magnon-electron scattering that leads to a finite interfacial spin conductance (per area) $g_s$. This yields
\begin{equation}
\label{eq:bcmagnonspincurrent}
\left. \hbar  {j}_m \right|_{\rm int}= g_s \left. \left(\mu_m - \mu_s^N\right) \right|_{\rm int}~.
\end{equation}
The interface spin conductance is proportional to the interface spin-mixing conductance $g_{\uparrow_\downarrow}$, and is given by
\begin{equation}
 g_s = \frac{3\zeta\left(\frac{3}{2}\right)g_{\uparrow\downarrow}}{2\pi s \Lambda^3}~,
\end{equation}
with $\Lambda = \sqrt{2 \pi \hbar^2/M k_B T}$ the magnon thermal deBroglie wavelength, $s$ the saturated spin density of the ferromagnet, and $\zeta (z)$ is the Riemann zeta function evaluated in $z$. The mixing conductance is a quantity that is known for most ferromagnet$|$normal-metal interfaces, either experimentally (see e.g. Ref.~\cite{PhysRevLett.111.176601}) or from ab initio computations \cite{jia2011}.

\section{Application}
\label{sec:appl}
\begin{figure}
	\includegraphics[width=15cm]{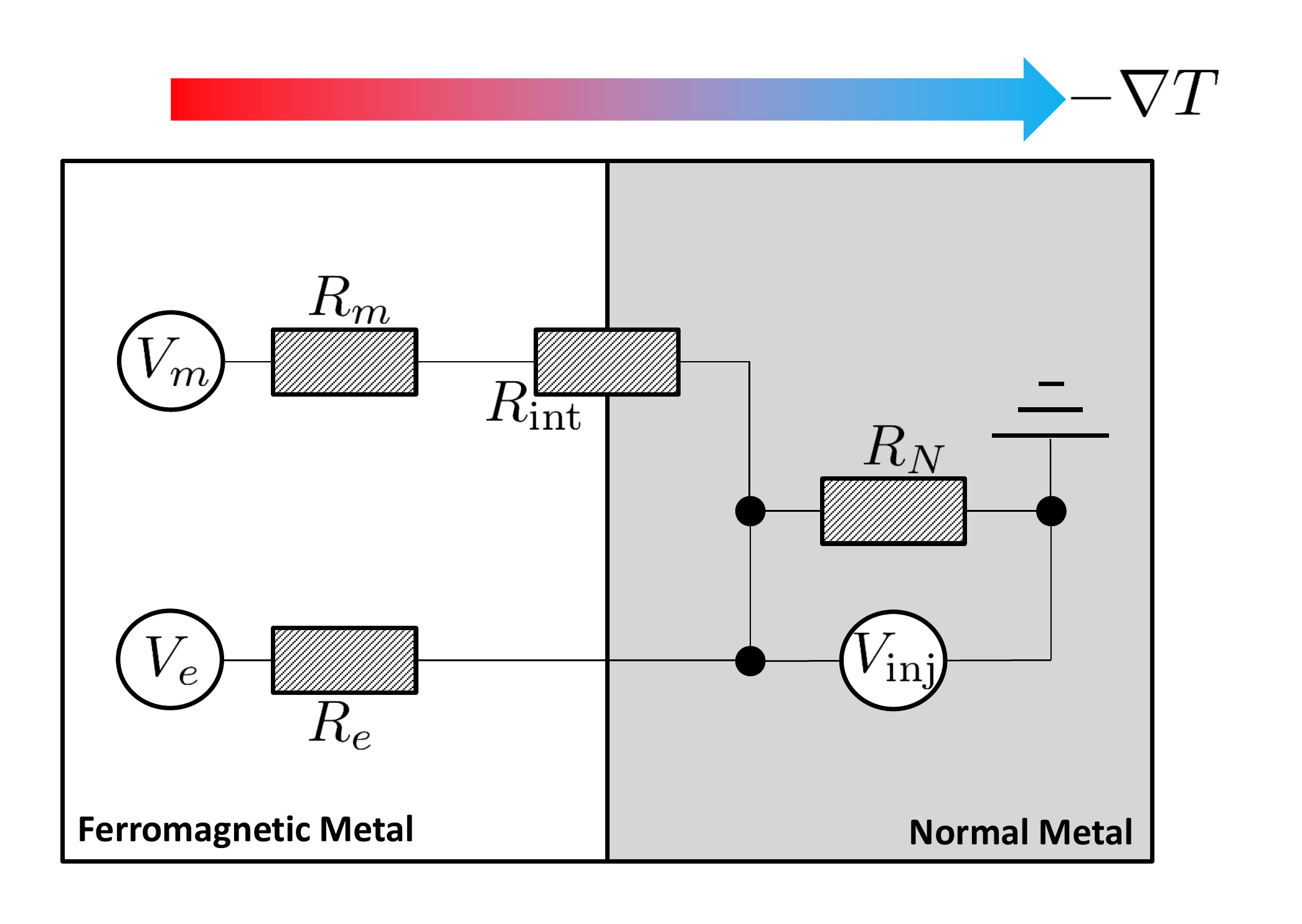} 
	\caption{Equivalent circuit for thermal spin injection across a ferromagnet$|$normal-metal interface.}
	\label{fig:circuit}
\end{figure}

As an application of our theory, we consider the set-up in Fig.~\ref{fig:system} and determine --- as a measure for the efficiency of the thermal spin injection --- the spin accumulation at the interface on the normal-metal side that results from the temperature gradient. For simplicity, we take this temperature gradient to be constant across the whole system, and, in particular, neglect interfacial Kapitza resistances.  We solve the equations for the magnon chemical potential and spin accumulation [Eqs.~(\ref{eq:steadystateeqschempots})] in the simplifying limit that $\ell_{\rm sd} \ll \ell_{\rm sf} \ll \ell_{\rm mr}$. The former of these is motivated by realizing the $\ell_{\rm sd}$ is limited by the non-relativistic $s$-$d$-exchange interactions, whereas $\ell_{\rm sf}$ and $\ell_{\rm mr}$ result from relativistic effects, i.e., spin-orbit coupling, which are typically weak. The limit $\ell_{\rm sf} \ll \ell_{\rm mr}$ follows from the assumption that the $s$-electron spin density relaxes faster than that of magnons \cite{PhysRevB.92.180412}. Within these assumptions, we have from Eqs.~(\ref{eq:steadystateeqschempots}) that in the ferromagnet $\mu_m = \mu_s \propto e^{x/\ell_{\rm sf}}$, where we took the interface to be the $y$-$z$-plane. In the normal metal we have, of course, no magnon chemical potential, and we have for the electronic spin accumulation that $\mu_s \propto e^{-x/\ell_{\rm sf}^N}$. 

Using the expressions for the currents in Eqs.~(\ref{eq:electronspincurrent}),~(\ref{eq:magnonspincurrent}),~and~(\ref{eq:electronspincurrentnormal}), together with the boundary conditions in Eqs.~(\ref{eq:bcspincurrents})~and~(\ref{eq:bcmagnonspincurrent}), and imposing that there is no charge current, we ultimately find that 
\begin{equation}
\label{resultspinaccumulation}
 \left. \mu_s^N \right|_{\rm int} = 
 \frac{e \ell_{\rm sf} \ell_{\rm sf}^N \nabla T \left[-2e\hbar  \ell_{\rm sf}g_sL +T\sigma_e (\sigma_m + g_s \ell_{\rm sf}\hbar )(S_\uparrow-S_\downarrow)\right]}
 {T\left[2e^2\ell_{\rm sf} \ell_{\rm sf}^N g_s \sigma_m +\hbar (\ell_{\rm sf}^N \sigma_e+\ell_{\rm sf}\sigma^N)(\sigma_m+g_s\ell_{\rm sf}\hbar)\right]}~,
\end{equation}
where we took $\sigma_\uparrow =\sigma_\downarrow \equiv \sigma_e$ to reduce the number of parameters. Moreover, the spin-dependence of the conductivities does not play an essential role in the discussion of the thermal spin injection as it is driven by the spin-dependence of the electronic Seebeck coefficients $S_\alpha$.

As a measure for the relative importance of the magnonic to the electronic contribution, we use the ratio between the first ($\propto L$) and second ($\propto (S_\uparrow-S_\downarrow)$) term in the result in Eq.~(\ref{resultspinaccumulation}). We use that $g_s,\sigma_m/\tau_{{\rm tr},m} \propto (T/T_C)^{3/2}$ and that $L/\tau_{{\rm tr},m}\propto T (T/T_C)^{3/2}$, where $T_C \sim \hbar^2 s^{2/3}/k_B M$. These temperature dependencies may be understood by noting that the magnon density scales as $(T/T_C)^{3/2}$. For the electron contribution we use that $S_\alpha \propto T/T_F$ for $T \ll T_F$. Hence, the ratio between the first  and second term of Eq.~(\ref{resultspinaccumulation}), corresponding respectively to the magnonic and electronic contribution, scales as $\sim \sqrt{T/T_C}T_F/T_C$. This implies that the magnonic contribution can be comparable or may even dominate over the electronic one. In materials where magnon drag is important we include the magnon-drag contribution to the electronic spin Seebeck coefficients, i.e., we take $S_\alpha \propto (T/T_C)^{3/2} $\cite{PhysRevB.94.144407}. This would lead to a ratio between magnonic and electronic contributions that is independent of temperature.

Note that the magnonic contribution is suppressed by the interface spin resistance for magnon spin currents that arises as a result of the interfacial $s$-$d$-scattering. This can be understood in terms of the equivalent circuit for the thermal spin transport sketched in Fig.~\ref{fig:circuit}. It relies on the fact that the finite spin relaxation lengths limit the spin accumulation drops to occur over equivalent resistances $R_m =\hbar^2 \ell_{\rm sf}/2A e^2 \sigma_m$ (with $A$ the cross section) for the magnon spin transport,  $R_e = \ell_{\rm sf}/A \sigma_e$ for the electron spin transport in the ferromagnet, and $R_{\rm int} = \hbar/2A e^2 g_s$ for the magnon spin current across the interface. Since we have taken the interface to be transparent for electrons, there is no interface spin resistance for the electron contribution. In the normal metal we have $R_N=\ell_{\rm sf}^N/(A \sigma^N)$. Furthermore, the spin transport is driven by a thermally-induced spin voltage $V_e=e \ell_{\rm sf} (S_\uparrow-S_\downarrow) \nabla T$ for the electrons and $V_m = -2 e^2 R_m L \nabla T/T$ for the magnons. Calculating $V_{\rm inj}$ from the equivalent circuit then  yields $V_{\rm inj} = \left. \mu_s^N \right|_{\rm int}/e$ and indeed reproduces Eq.~(\ref{resultspinaccumulation}). Using the equivalent circuit, it is clear that, because the magnon and electron contribution occur in parallel, their relative contribution picks up the factor $R_e/(R_m+R_{\rm int})$. Using our previous estimates we find that 
\begin{equation}
 \left(\frac{R_e}{R_m+R_{\rm int}} \right) \sim \left(\frac{T}{T_C}\right)^{3/2} \left(\frac{1}{{\frac{M\tau_{{\rm tr},e}}{m \tau_{\rm sd}}} + \frac{k_F^2 \tau_{{\rm tr},e}}{g_{\uparrow\downarrow}\tau_{\rm sf}}}\right) \sim \left(\frac{T}{T_C}\right)^{3/2}~,
\end{equation}
where $k_F$ is the Fermi wave vector, and where we took $\ell_{\rm sf} \sim \hbar k_F \tau_{\rm sf}/m$ and $n_e \sim k_F^3$, and used that typically $g_{\uparrow\downarrow} \sim 1/k_F^2$, $\tau_{\rm sf} \gg \tau_{{\rm tr},e}$, and took, using the values estimated before, that  $M\tau_{{\rm tr},e}/m \tau_{\rm sd}\sim 1$.

\section{Discussion, conclusions and outlook}
\label{sec:discconcloutl}
In conclusion, we have developed a theory for thermal spin injection from a ferromagnetic metal into a normal metal that takes magnons and electrons into account on equal footing, and have shown that the magnon contribution can in general not be neglected with respect to the electronic one. While we have made various simplifying assumptions along the way, this main conclusion is not affected by these assumptions, as it ultimately relies on the fact that the scales for thermal transport by magnons and electrons are set by the Curie temperature and the Fermi temperature, respectively.  

A useful direction for future research is to analyze the experiments on the spin-dependent Seebeck effect \cite{slachter2010} in detail starting from the framework developed here. Our theory is also natural starting point for the inclusion of transport processes in ultrafast magnetization dynamics as described by e.g. the microscopic three-temperature model \cite{PhysRevLett.95.267207}.

YT and JPH acknowledge support from the Center for Emergent Materials: an NSF MRSEC under award number DMR-1420451. RD is a member of the D-ITP consortium, a program of the Netherlands Organisation for Scientific Research (NWO) that is funded by the Dutch Ministry of Education, Culture and Science (OCW). This work is in part funded by the Stichting voor Fundamenteel Onderzoek der Materie (FOM) and the European Research Council (ERC).

\section*{References}
\providecommand{\newblock}{}

\end{document}